\documentclass[12pt]{article}

\usepackage{graphicx}
\usepackage{amssymb}
\begin{document}

\newcommand{\vecx}{\mbox{\boldmath $x$}}


%
\begin{center}
{\large\bf
The interpolation approach to \\
nonextensive quantum systems 
} 
\end{center}

\begin{center}
Hideo Hasegawa
\footnote{hideohasegawa@goo.jp}
\end{center}

\begin{center}
{\it Department of Physics, Tokyo Gakugei University,  \\
Koganei, Tokyo 184-8501, Japan}
\end{center}
\begin{center}
({\today})
\end{center}
\thispagestyle{myheadings}

\begin{abstract}
Recently it has been shown by the present author
[H. Hasegawa, Phys. Rev. E (in press): arXiv:0904.2399] 
that the interpolation approximation (IA) to
the generalized Bose-Einstein and Femi-Dirac distributions   
yields results in agreement with the exact ones within
the $O(q-1)$ and in high- and low-temperature limits,
where $(q-1)$ expresses the non-extensivity:
the case of $q=1$ corresponding to the conventional quantal distributions.
In this study, we have applied the generalized distributions in the IA 
to typical nonextensive subjects: (1) the black-body radiation,
(2) the Bose-Einstein condensation, (3) the BCS superconductivity and
(4) itinerant-electron (metallic) ferromagnetism.
Effects of the non-extensivity on physical quantities in these
nonextenisive quantum systems have been investigated.
A critical comparison is made between 
results calculated by the IA and the factorization approximation (FA)
which has been so far applied to many nonextensive systems.
It has been pointed out that the FA overestimates the non-extensivity and that
it leads to an inappropriate results for fermion systems
like the subjects (3) and (4).

\end{abstract}

\noindent
\vspace{0.5cm}

{\it PACS No.}: 05.30.-d, 71.28.+d, 75.10.Lp, 75.40.Cx
\vspace{1cm}

\noindent

{\it Keywords}: nonextensive quantum statistics, black-body radiation, 
Bose-Einstein condensation, BCS superconductivity, itinerant-electron ferromagnetism
 

\newpage

\section{Introduction}

Considerable works have been made on the nonextensive statistics
since Tsallis proposed the generalized entropy (called the Tsallis entropy) 
\cite{Tsallis88} which is a one-parameter generalization of the Boltzmann-Gibbs 
entropy with the entropic index $q$: the Tsallis entropy
in the limit of $q=1.0$ reduces to the Boltzmann-Gibbs entropy
(for a recent review, see \cite{Tsallis04}). 
In recent years, much attention has been paid to an application
of the nonextensive statistics to quantum subjects,
in which the generalized Bose-Einstein and Fermi-Dirac 
distributions (called $q$-BED and $q$-FDD hereafter)
play important roles. The four methods have been proposed for
$q$-BED and $q$-FDD: 
(i) the asymptotic approach (AA) \cite{Tsallis95}
obtained the canonical partition function valid
within $ O((q-1)/k_B T)$,
(ii) the factorization approach (FA) \cite{Buy95} 
employed the decoupling, factorization approximation
in evaluating the grand-canonical partition function,
(iii) the exact approach (EA) \cite{Rajagopal98,Lenzi99} 
derived the formally exact expression 
for the grand canonical partition function expressed 
in terms of the Boltzmann-Gibbs counterpart, 
and (iv) the interpolation approximation (IA) \cite{Hasegawa09b}  
was proposed based on the EA, yielding results in agreement
with those obtained by the EA within $O(q-1)$ and 
in high- and low-temperature limits.
Among the four methods,
the FA has been mostly adopted in many quantum subjects
including the black-body radiation \cite{Tirnakli00,Tirnakli97,Wang98},
early universe \cite{Tirnakli99,Pessah01},
the Bose-Einstein condensation (BEC) \cite{Torres98}-\cite{Lawani08},
metals \cite{Oliveira00}, superconductivity \cite{Nunes01,Uys01},
spin systems \cite{Portesi95}-\cite{Reis06}
and itinerant-electron (metallic) ferromagnets \cite{Hasegawa09a}. 
This is due to a simplicity of the expression
of the generalized quantal distributions in the FA.

Quite recently, however, it has been pointed out 
that the FA is not accurate from a study of the EA \cite{Hasegawa09b}.
It might be necessary to examine calculations
previously made with the use of the FA by a new calculation 
with the IA, which is the purpose of the present paper.
We will discuss the four quantum subjects: (1) the black-body radiation,
(2) the Bose-Einstein condensation, (3) the BCS superconductivity and 
(4) the itinerant-electron ferromagnets, to which the FA has been applied
\cite{Tirnakli00,Tirnakli97,Wang98}\cite{Torres98}-\cite{Lawani08}
\cite{Nunes01,Uys01}\cite{Hasegawa09a}.

The paper is organized as follows. In Sec. 2, we 
briefly discuss the $q$-BED and $q$-FDD in the EA and IA 
after Ref. \cite{Hasegawa09b}.
Then we apply the IA to the four subjects mentioned above.
The black-body radiation is discussed in Sec. 3, where
the $q$-dependent Stefan-Boltzmann coefficient and
the Wien shift law are calculated.
In Sec. 4, we investigate the BEC, calculating the critical
temperature and the temperature dependence 
of the energy and specific heat as a function of $q$.
The $q$-FDD in the IA is applied to the BCS
superconductivity in Sec. 5, where the $q$ dependence of
the superconducting critical temperatures and the
characteristic ratios related with the ground-state order parameter,
critical temperature and specific heat in the BCS theory. 
In Sec. 6, we discuss magnetic and thermodynamical properties
of the itinerant-electron ferromagnets described
by the Hubbard model combined with the Hartree-Fock approximation.
In Sec. 7, we present qualitative discussions 
with the use of the generalized Sommerfeld low-temperature expansion.
Section 7 is devoted to our conclusion. 

\section{Generalized quantal distributions}

\subsection{An exact approach}

When the maximum entropy method with 
the optimal Lagrange multiplier \cite{Martinez00} is adopted,
the generalized quantal distribution for the state $k$ (whose number operator
is $\hat{n}_k$) is given by \cite{Hasegawa09b}
\begin{eqnarray}
f_q(\epsilon_k, \beta) 
&=& \frac{1}{X_q} 
Tr \{ [1+(q-1) \beta(\hat{H}-\mu \hat{N} - E_q + \mu N_q)]^{\frac{q}{1-q}}\:\hat{n}_k \},
\label{eq:M1} \\
&=& \frac{1}{X_q} \int_0^{\infty}
G\left(u;\frac{q}{q-1}, \frac{1}{(q-1)\beta} \right) \: 
e^{u(E_q-\mu N_q)} \:\Xi_1(u)
f_1(\epsilon_k,u)\:du \nonumber  \\
&& \label{eq:M2}
\hspace{8cm}\mbox{for $q > 1$}, \\
&=& \frac{i}{2 \pi X_q} 
\int_C  H\left(t;\frac{q}{1-q},\frac{1}{(1-q)\beta} \right)
e^{-t(E_q-\mu N_q)} \; \Xi_1(-t)
\:f_1(\epsilon_k,- t)\: dt \nonumber \\
&& \label{eq:M3}
\hspace{8cm}\mbox{for $q < 1$},
\end{eqnarray}
with
\begin{eqnarray}
X_q &=&
Tr \{ [1+(q-1) \beta(\hat{H}-\mu \hat{N} - E_q + \mu N_q)]^{\frac{1}{1-q}} \}, 
\label{eq:M4}\\
&=& \int_0^{\infty} 
G\left(u;\frac{1}{q-1}, \frac{1}{(q-1)\beta} \right) \: 
e^{u(E_q-\mu N_q)} \:\Xi_1(u) \:du  
\hspace{1cm}\mbox{for $q > 1$},
\label{eq:M5} \\
&=& \frac{i}{2 \pi } 
\int_C H\left(t;\frac{1}{1-q},\frac{1}{(1-q)\beta} \right)
e^{-t(E_q-\mu N_q)} \; \Xi_1(-t) \:dt
\hspace{1cm}\mbox{for $q < 1$}, 
\label{eq:M6}
\end{eqnarray}
where
\begin{eqnarray}
\Xi_1(u) &=& e^{- u \:\Omega_1(u)}
=Tr \{e^{-u (\hat{H}-\mu \hat{N})} \}
= \prod_k [1 \mp e^{-u(\epsilon_k-\mu)}]^{\mp 1}, 
\label{eq:M9} \\
\Omega_1(u) &=& 
\pm \frac{1}{u} \sum_k \ln[1 \mp e^{-u(\epsilon_k-\mu)}], 
\label{eq:M10} \\
%
%
%
f_1(\epsilon,u) &=& \frac{1}{e^{u(\epsilon-\mu)} \mp 1},
\label{eq:M7} \\
G\left(u;a,b \right) 
&=& \frac{b^a}{\Gamma\left(a  \right)} 
u^{a-1} e^{-b u},
\label{eq:M8} \\
H(t;a,b)&=&  \: \Gamma(a+1) b^{-a} \:(-t)^{-a-1} e^{- b t}.
\end{eqnarray}
Upper and lower signs in Eqs. (\ref{eq:M9})-(\ref{eq:M7})
are applied to boson and fermion, respectively,
$\Gamma(z)$ stands for the gamma function and
$C$ denotes the Hankel path in the complex plane \cite{Rajagopal98,Lenzi99}.
It is noted that Eqs. (\ref{eq:M1})-(\ref{eq:M6}) 
include $N_q$ and $E_q$ which should be
determined in a self-consistent way
[see Eqs. (2) and (3) in Ref. \cite{Hasegawa09b}].
Such self-consistent calculations have been reported for the band electron model 
and the Debye phonon model in Ref. \cite{Hasegawa09b}.

\subsection{The interpolation approximation}

Self-consistent calculations for $f_q(\epsilon_k, \beta)$
including $N_q$ and $E_q$ are rather tedious and difficult.
In order to overcome this difficulty, we have proposed
the IA \cite{Hasegawa09b}, assuming 
that
\begin{eqnarray}
\frac{1}{X_q} \:e^{u(E_q-\mu N_q)} \:\Xi_1(u) &=& 1,
\end{eqnarray}
in Eqs. (\ref{eq:M2}) and (\ref{eq:M3}).
Then the generalized distribution in the IA is given by 
\begin{eqnarray}
f_q^{IA}(\epsilon, \beta)
&=& 
\frac{1}{\Gamma(\frac{q}{q-1})} \left( \frac{1}{(q-1)\beta} \right)^{\frac{q}{q-1}}
\int_0^{\infty} u^{\frac{1}{q-1}} e^{-\frac{u}{(q-1)\beta}}
\:f_1(\epsilon, u)\:du \nonumber \\
&& \hspace{8cm} \mbox{for $q > 1.0$},
\label{eq:N1} \\
%
&=& \frac{\Gamma\left(\frac{1}{1-q} \right)}{[(1-q)\beta]^{-\frac{q}{1-q}}} 
\left(\frac{i}{2 \pi} \right) 
\int_C\:(-t)^{-\frac{1}{1-q}} e^{-\frac{t}{(1-q)\beta}}
\:f_1(\epsilon, -t)\: dt \nonumber \\
&&\hspace{8cm} \mbox{for $q < 1.0$}. 
\label{eq:N2} 
\end{eqnarray}

\noindent
{\bf $q$-BED}

With the use of Eqs. (\ref{eq:N1}) and (\ref{eq:N2}),
the analytic expression of the $q$-BED in the IA is given by \cite{Hasegawa09b}
\begin{eqnarray}
f_q^{IA}(\epsilon,\beta) 
&=& \sum_{n=0}^{\infty}\:[e_q^{-(n+1)\:x}]^q
\hspace{3cm}\mbox{for $0 < q < 3 $}, 
\label{eq:N3}
\end{eqnarray}
where $e_q^x$ expresses the $q$-exponential function defined by
\begin{eqnarray}
e_q^{x} &=& \exp_q(x) = [1+(1-q)x]^{\frac{1}{1-q}}
\hspace{1cm}\mbox{for $1+(1-q)x > 0$}, \\
&=& 0
\hspace{5.5cm}\mbox{for $1+(1-q)x \leq 0$},
\label{eq:N4}
\end{eqnarray}
with the cut-off properties.

\noindent
{\bf $q$-FDD}

Similarly, the analytic expression of the $q$-FDD in the IA 
is given by \cite{Hasegawa09b}
\begin{eqnarray}
f_q^{IA}(\epsilon, \beta) 
&=& F(\epsilon, \beta)
\hspace{4cm} \mbox{for $ \epsilon  > \mu$},
\label{eq:N5} \\
&=& \frac{1}{2}
\hspace{5cm} \mbox{for $ \epsilon  = \mu$}, \\
&=& 1.0 - F(\vert \epsilon -\mu \vert +\mu, \beta)
\hspace{1cm} \mbox{for $ \epsilon < \mu $},
\label{eq:N6}
\end{eqnarray}
with
\begin{eqnarray}
F(\epsilon,\beta) &=& 
\sum_{n=0}^{\infty} \: \: (-1)^n [e_q^{-(n+1) x}]^q
\hspace{2.5cm}\mbox{for $0 < q < 3$}. 
\label{eq:N7} 
\end{eqnarray}
Note that 
$e_q^{-(n+1) x}=[1-(1-q)(n+1)x]^{\frac{1}{(1-q)}} \neq [e_q^{-x}]^{(n+1)}$ 
in Eqs. (\ref{eq:N3}) and (\ref{eq:N7}) except for $q=1.0$.
$f_q^{IA}(\epsilon, \beta)$ given by Eqs. (\ref{eq:N3})-(\ref{eq:N7}) 
reduces to $f_1(\epsilon, \beta)$ in the limit of $q \rightarrow 1.0$
where $e_q^{x} \rightarrow e^{x}$. 

On the contrary, the $q$-BED and $q$-FDD in the FA are given by \cite{Buy95}
\begin{eqnarray}
f_q^{FA}(\epsilon,\beta) &=& 
\frac{1}{(e_q^{-x})^{-1} \mp 1}
= \sum_{n=0}^{\infty} (e_q^{-x})^{n+1},
\label{eq:N8}
\end{eqnarray}
where the upper (lower) sign is applied to boson (fermion).
It is noted that if we adopt a factorization approximation:
$e_q^{-(n+1)x} \simeq [e_q^{-x}]^{(n+1)}$ 
in Eqs. (\ref{eq:N3})-(\ref{eq:N7}),
we obtain
\begin{eqnarray}
f_q^{FAq}(\epsilon,\beta) &\simeq & 
\sum_{n=0}^{\infty} (e_q^{-x})^{(n+1)q}
=\frac{1}{ (e_q^{-x})^{-q} \mp 1}, 
\label{eq:N9}
\end{eqnarray} 
which is similar to Eq. (\ref{eq:N8})
and which is referred to as the FAq hereafter.

A comparison among the $O(q-1)$ contributions to
the generalized quantal distributions in the EA, IA, FA and FAq is made in Table 1.
It is stressed that the IA has the interpolation character
yielding good results in the limits of $q \rightarrow 1.0$,
$\beta \rightarrow \infty$ and $\beta \rightarrow 0.0$
\cite{Hasegawa09b}.
The $O(q-1)$ contributions in the FA and FAq are different from that in the EA.
In the limit of $\beta \rightarrow 0.0$, $f_q^{EA}(\epsilon)$,
$f_q^{IA}(\epsilon)$ and $f_q^{FAq}(\epsilon)$
reduce to $[e_q^{-\beta \epsilon}]^q$, while
$f_q^{FA}$ reduces to $e_q^{-\beta \epsilon}$.
In the limit of $\beta \rightarrow \infty$, all the $q$-FDD 
become $\Theta(\mu-\epsilon)$, where
$\Theta(x)$ denotes the Heaviside function.
More detailed comparisons among various methods
have been made in Ref.\cite{Hasegawa09b}.

\begin{table}
\begin{center}
\caption{Generalized quantal distributions in the limits of 
$q \rightarrow 1$, $\beta \rightarrow \infty$ and  $\beta \rightarrow 0$}
\vspace{0.5cm}
\renewcommand{\arraystretch}{2.5}
\begin{tabular}{|c|c|c|c|} \hline
method &  $q \rightarrow 1 $ & $\beta \rightarrow \infty$ (FDD) 
&  $\beta \rightarrow 0$ \\ \hline \hline
${\rm EA}^{a}$  & $f_1
+ (q-1)\left[(\epsilon-\mu) \:\frac{\partial f_1}{\partial \epsilon}
+ \frac{1}{2}(\epsilon-\mu)^2 \:\frac{\partial^2 f_1}{\partial \epsilon^2} \right]$  
&  $\Theta(\mu-\epsilon)$  &  $[e_q^{-\beta(\epsilon-\mu)}]^q$ \\ \hline
${\rm IA}^{b}$  & $f_1
+ (q-1)\left[(\epsilon-\mu) \:\frac{\partial f_1}{\partial \epsilon}
+ \frac{1}{2}(\epsilon-\mu)^2 \:\frac{\partial^2 f_1}{\partial \epsilon^2} \right]$  
&  $\Theta(\mu-\epsilon)$  & $[e_q^{-\beta(\epsilon-\mu)}]^q$  \\ \hline
${\rm FA}^{c}$  & $f_1 - \frac{1}{2}(q-1) \beta (\epsilon-\mu)^2
\: \frac{\partial f_1}{\partial \epsilon} $ 
& $\Theta(\mu-\epsilon)$  & $ e_q^{-\beta(\epsilon-\mu)}$ \\ \hline
${\rm FAq}^{d}$  & $f_1 
+(q-1)[(\epsilon-\mu)- \frac{1}{2}\beta (\epsilon-\mu)^2]
\: \frac{\partial f_1}{\partial \epsilon} $ 
& $\Theta(\mu-\epsilon)$  & $ [e_q^{-\beta(\epsilon-\mu)}]^q$ \\ \hline
\end{tabular}
\end{center}

\noindent 
$f_1=1/(e^{\beta(\epsilon-\mu)} \mp 1)$: 
$\Theta(x)$, the Heaviside function: 
$e_q^{x}$, $q$-exponential function.

\noindent
$^a$ the exact approach \cite{Hasegawa09b}

\noindent
$^b$ the interpolation approximation \cite{Hasegawa09b}

\noindent
$^c$ the factorization approximation \cite{Buy95}

\noindent
$^d$ the factorization approximation [Eq. (\ref{eq:N9})]

\end{table}

\section{Black-body radiation}

We will apply the $q$-BED given by Eq. (\ref{eq:N3}) in the IA to 
the black-body radiation model with the photon density of states 
per volume given by
\begin{eqnarray}
\rho(\omega) &=& C \omega^2,
\end{eqnarray}
where $C=1/\pi^2 c^3$ and $c$ denotes the light velocity.
The generalized Planck law is given by
\begin{eqnarray}
D_q(\omega) &=& \hbar \omega \:\rho(\omega) f_q^{IA}(\hbar \omega,\beta).
\label{eq:R1}
\end{eqnarray}
The $q$-BED in the FA was adopted to the black-body radiation
in Refs. \cite{Tirnakli00,Tirnakli97,Wang98}.

The $q$-BEDs (with $\mu=0.0$) calculated in the IA and FA 
are shown by solid and dashed curves, respectively, in Fig. \ref{figL}(a)
with the logarithmic ordinate: they are indistinguishable 
in the linear scale.
For $q=1.2$, tails of $q$-BED obey the power law. In contrast,
for $q=0.8$, $q$-BED has a compact form with the cut-off behavior:
$f_q(\hbar \omega)=0.0$ for $\beta \hbar \omega \geq 5.0$.
Solid and dashed curve in Fig. \ref{figL}(b) express 
the generalized Planck law $D_q(\omega)$
calculated by the IA and FA, respectively.
For $q=1.2$, the distribution of
$D_q(\omega)$ in the high-frequency region is much increased.
This trend is reversed for $q=0.8$.
The effect of the non-extensivity in the FA is much 
overestimated compared to that in the IA. 

We obtain the generalized Stefan-Boltzmann law,
\begin{eqnarray} 
E_q &=& \int_0^{\infty}\: D_q(\omega) \:d \omega, \\
&=& \sigma_q \:T^4,
\label{eq:R2}
\end{eqnarray}
with
\begin{eqnarray}
\frac{\sigma_q}{\sigma_1} &=& \frac{\Gamma(\frac{1}{q-1}-3)}
{(q-1)^4 \Gamma(\frac{1}{q-1}+1)} 
\hspace{1cm}\mbox{for $q > 1.0$},
\label{eq:R3} \\
&=& \frac{\Gamma(\frac{q}{1-q}+1)}
{(1-q)^4 \Gamma(\frac{q}{1-q}+5)} 
\hspace{1cm}\mbox{for $q > 1.0$}, 
\label{eq:R4} \\
&=& \frac{1}{(2-q)(3-2 q)(4-3 q)}
\hspace{1cm}\mbox{for $0 <q < 4/3$},
\label{eq:R5}
\end{eqnarray}
where $\sigma_1$ is the Stefan-Boltzmann constant for $q=1.0$.
The $q$ dependence of $\sigma_q$ calculated in the IA and FA
are shown by solid and dashed curves, respectively,
in Fig. \ref{figK} with the logarithmic ordinate.
With increasing $q$, $\sigma$ is monotonously increased.

Substituting Eq. (\ref{eq:N3}) to Eq. (\ref{eq:R1}), 
we obtain $\omega_{m}$ where $D_q(\omega,\beta)$ 
has the maximum,
\begin{eqnarray}
\omega_{m} &=& \left (\frac{3 f_q^{IA}(\hbar \omega,\beta)}
{[-\frac{\partial}{\partial \omega} f_q^{IA}(\hbar \omega,\beta)]}
\right)_{\omega=\omega_m}, \\
&=& \left( \frac{3 k_B T}{\hbar} \right)
\frac{\sum_{n=0}^{\infty} \: \frac{1}
{n!} [e_q^{-(n+1) \beta \hbar \omega_{m}}]^{q} }
{q \sum_{n=0}^{\infty} \: \frac{(n+1)}
{n!}[e_q^{-(n+1) \beta \hbar \omega_{m} }]^{(2q-1)} }, \\
&\rightarrow& \left( \frac{3 k_B T}{\hbar} \right)
(1-e^{-\beta \hbar \omega_m})
\hspace{1cm}\mbox{for $q \rightarrow 1.0$},
\end{eqnarray}
whose solution expresses the generalized Wien shift law.
The solid curve in Fig. \ref{figJ} shows the calculated ratio of 
$\omega_{m,q}/\omega_{m,1}$ as a function of $q$.
With increasing $q$ above $q=1.0$, the ratio is increased whereas
it is decreased with decreasing $q$ below unity.
Dashed and chain curves show the results of the FA
and AA [$\omega_{m,q}/\omega_{m,1} = 1+6.16 \:(q-1)$] \cite{Tsallis95},
respectively.

\section{Bose-Einstein condensation} 

\subsection{Basic equation}

We will study this subject, applying the IA
to a bose gas with the density of states given by
\begin{eqnarray}
\rho(\epsilon) &=& A \:\epsilon^{r},
\label{eq:P1}
\end{eqnarray}
where $r=d/2-1$ for $d$-dimensional ideal bose gase, 
$r=d-1$ for bose gase trapped in $d$-dimensional harmonic potential,
and $A$ stands for a relevant coefficient.
The nonextensive BEC was discussed in Refs.
\cite{Torres98}-\cite{Lawani08} with the use of the FA.

By using the $q$-BED given by Eq. (\ref{eq:N3}) 
in the IA, we obtain the number of electrons given by
\begin{eqnarray}
N &=& N_c+N_e,
\end{eqnarray}
with
\begin{eqnarray}
N_c &=& \frac{1}{e^{\alpha}-1}=\sum_{n=0}^{\infty}\: e^{-(n+1) \alpha},
\hspace{1cm}\mbox{for $q =1$}, \\
&=& \sum_{n=0}^{\infty} \:[e_q^{-(n+1) \alpha}]^q
\hspace{2cm}\mbox{for $0 < q < 3$},
\label{eq:P2} \\ 
N_e &=& (k_B T)^{r+1} 
\frac{A \:\Gamma(r+1)\Gamma(\frac{1}{q-1}-r)}
{(q-1)^{r+1} \Gamma(\frac{1}{q-1}+1)} 
\phi_q(r+1, \alpha)
\hspace{0.5cm}\mbox{for $1 < q < 3 $}, 
\label{eq:P3} \\
&= & (k_B T)^{r+1} A \:\Gamma\left(r+1 \right)  \phi(r+1, \alpha)
\hspace{2cm}\mbox{for $ q = 1$}, \\
&=& (k_B T)^{r+1} \frac{A \:\Gamma(r+1)\Gamma(\frac{q}{1-q}+1)}
{(1-q)^{r+1} \Gamma(\frac{q}{1-q}+r+2)}
\phi_q(r+1, \alpha)
\hspace{0.2cm}\mbox{for $0< q < 1$},
\label{eq:P4}
\end{eqnarray}
where $\alpha=- \beta \mu$ ($\geq 0$), and
$N_c$ and $N_e$ denote the numbers of electrons
in the condensed and excited states, respectively.
Here $\phi_q(z,\alpha)$ is the generalized
Bose integral defined by
\begin{eqnarray}
\phi_q(z, \alpha) &\equiv & \sum_{n=1}^{\infty}
\frac{[e_q^{-n \alpha }]^{z-(z-1)q}}{n^z}
\hspace{1cm}\mbox{for $ \Re z > 1 $},
\label{eq:P5}
\end{eqnarray}
which reduces to
\begin{eqnarray}
\phi_q(z,\alpha)
&\rightarrow& \sum_{n=1}^{\infty} \frac{1}{n^z}
= \zeta(z)
\hspace{2cm}\mbox{for $\alpha \rightarrow 0.0$}, \\
&\rightarrow& \sum_{n=1}^{\infty} \frac{ e^{-n \alpha }}{n^z}
= \phi(z,\alpha)
\hspace{1cm}\mbox{for $q \rightarrow 1.0$}, 
\end{eqnarray}
$\zeta(z)$ and $\phi(z,\alpha)$  being the Riemann zeta function and
the Bose integral, respectively.

\subsection{Critical temperature}

The number of electrons in the excited state is bounded by
Eqs. (\ref{eq:P3})-(\ref{eq:P4}) with $\alpha=0.0$.
Then the critical temperature of the BEC, $T_c$,  
below which $\alpha$ vanishes is given by
\begin{eqnarray}
k_B T_{c} &=& (q-1) \left[\frac{N \Gamma(\frac{1}{q-1}+1)}
{A \Gamma(r+1) \zeta(r+1) 
\Gamma(\frac{1}{q-1}-r)} \right]^{\frac{1}{r+1}}
\hspace{1cm}\mbox{for $1 < q < 3$}, 
\label{eq:P6} \\
&=& \left[\frac{N}{A \Gamma(r+1)\zeta(r+1)}\right]^{\frac{1}{r+1}}
\hspace{1cm}\mbox{for $q = 1$}, 
\label{eq:P7} \\
&=& (1-q) \left[\frac{N \Gamma(\frac{q}{1-q}+r+2)}
{A \Gamma(r+1) \zeta(r+1) 
\Gamma(\frac{q}{1-q}+1)} \right]^{\frac{1}{r+1}}
\hspace{1cm}\mbox{for $0< q < 1$},
\label{eq:P8}
\end{eqnarray}
We note that $T_{c}=0$ for $r=0$ ({\it i.e.,} free boson with $d=2$
or $d=1$ boson with harmonic-potential)
because $\zeta(1)=\infty$.
Equations (\ref{eq:P6})-(\ref{eq:P8}) lead to
\begin{eqnarray}
\frac{T_{c,q}}{T_{c,1}} &=& (q-1) 
\left[\frac{\Gamma(\frac{1}{q-1}+1)}
{\Gamma(\frac{1}{q-1}-r)} \right]^{\frac{1}{r+1}}
\hspace{1.5cm}\mbox{for $1 < q < 3$}, 
\label{eq:Q6}\\
&=& (1-q) 
\left[\frac{\Gamma(\frac{q}{1-q}+r+2)}
{\Gamma(\frac{q}{1-q}+1)} \right]^{\frac{1}{r+1}}
\hspace{1cm}\mbox{for $0 < q < 1$}.
\label{eq:Q7} 
\end{eqnarray}
The solid curve in Figs. \ref{figG} (a) and (b) show the 
$q$ dependence of the ratio of $T_{c,q}/T_{c,1}$ for $r=1/2$
and $r=2$, respectively, 
calculated by Eqs. (\ref{eq:Q6}) and (\ref{eq:Q7}). 
The critical temperature is decreased with increasing $q$. 

On the other hand, the critical temperature in the FA is given by
\cite{Ou05} 
\begin{eqnarray}
\frac{T_{c,q}}{T_{c,1}} 
&=& \left[ \frac{\zeta(r+1)}{\zeta_q(r+1)} \right]^{\frac{1}{r+1}},
\label{eq:Q8}
\end{eqnarray}
with 
\begin{eqnarray}
\zeta_q(r+1) 
&=& \frac{1}{(q-1)^{r+1}}\sum_{n=1}^{\infty} 
\:\frac{\Gamma(\frac{n}{q-1}-r-1)}
{\Gamma(\frac{n}{q-1})}
\hspace{1cm}\mbox{for $1 < q <3$}, \\
&=& \zeta(r+1)
\hspace{5cm}\mbox{for $q=1$}, \\
&=& \frac{1}{(1-q)^{r+1}}\sum_{n=1}^{\infty} 
\:\frac{\Gamma(\frac{n}{1-q}+1)}
{\Gamma(\frac{n}{1-q}+r+2)}
\hspace{1cm}\mbox{for $0 < q < 1$}.
\label{eq:Q9} 
\end{eqnarray}
Dashed curves in Figs. \ref{figG} (a) and (b)
express the results of the FA calculated by Eqs. (\ref{eq:Q8})-(\ref{eq:Q9}).
The effect of the non-extensivity is overestimated in the FA:
$T_c^{FA}$ vanishes at $q \geq 1.67$
and $q \geq 1.33$ for $r=0.5$ and $r=2.0$, respectively.
In contrast, $T_c^{IA}$ vanishes at $q \geq 3.0$
and $q \geq 1.5$ for $r=0.5$ and $r=2.0$, respectively.

\subsection{Condensed states}

The number of electrons in the condensed states $N_c$ is
given by
\begin{eqnarray}
\frac{N_c}{N} &=& 1 - \left( \frac{T}{T_c}\right)^{r+1}
\hspace{1cm}\mbox{for $T \leq T_c$},
\end{eqnarray}
which depends on $r$ but it is independent of $q$.

\subsection{Energy and specific heat}

The total energy is given by
\begin{eqnarray}
E &=& (k_B T)^{r+2} \frac{A \:\Gamma(r+2)\Gamma(\frac{1}{q-1}-r-1)}
{(q-1)^{r+2} \Gamma(\frac{1}{q-1}+1)}
\phi_q(r+2, \alpha)
\hspace{0.5cm}\mbox{for $q \neq 1 $}, 
\label{eq:S1} \\
&=& (k_B T)^{r+2} A \:\Gamma\left(r+2 \right)  \phi(r+2, e^{-\alpha})
\hspace{3cm}\mbox{for $ q = 1$}.
\label{eq:S2}
\end{eqnarray}

Above $T_c$, $\alpha$ is temperature dependent because
it is adjusted as to conserve the total number
of electrons,
\begin{eqnarray}
1 &=& \left(\frac{T}{T_c}\right)^{r+1} \frac{\phi_q(r+1, \alpha)}{ \zeta(r+1)}.
\end{eqnarray}
Then its temperature dependence is given by
\begin{eqnarray}
\frac{d \alpha}{d T} &=& \frac{(r+1)}{(r+1-r q)\: T}
\frac{\phi_q(r+1,\alpha)}{\phi_q(r,\alpha)}
\hspace{1cm}\mbox{for $ q \neq 1$}, \\
&=&\frac{(r+1) \: \phi(r+1, \alpha)}
{T \:\phi(r, \alpha)}
\hspace{2cm}\mbox{for $ q = 1$}.
\end{eqnarray}
Taking into account the temperature dependence of $\alpha$,
we obtain the specific heat at $T \geq T_c$ given by
\begin{eqnarray}
\frac{C}{k_B N} 
&=& \left( \frac{T}{T_c}\right)^{r+1}  
\frac{(r+1)}
{[r+2-(r+1) q] \zeta(r+1)}
\{ (r+2) \phi_q(r+2,\alpha) \nonumber \\
&& -\frac{(r+1)[r+2-(r+1)q] [\phi_q(r+1, \alpha)]^2}
{\phi_q(r, \alpha)} \}
\hspace{0.5cm}\mbox{for $ q \neq 1$}, 
\label{eq:S3} \\
&=& \left(\frac{T}{T_c}\right)^{r+1} \frac{(r+1)}{\zeta(r+1)}
\left[ (r+2) \phi(r+2, \alpha) -(r+1) 
\frac{[\phi(r+1, \alpha)]^2}{\phi(r, \alpha)} \right] \nonumber \\
&& \hspace{8cm}\mbox{for $ q = 1$}.
\label{eq:S4}
\end{eqnarray}
Below $T_c$ where $\alpha=0.0$, we obtain the specific heat
given by
\begin{eqnarray}
\frac{C}{k_B N} &=& \left( \frac{T}{T_c}\right)^{r+1}  
\frac{(r+1) (r+2)\zeta(r+2)}
{[r+2-(r+1)q] \zeta(r+1)}
\hspace{0.5cm}\mbox{for $ q \neq 1$}, 
\label{eq:S5} \\
&=&\left( \frac{T}{T_c}\right)^{r+1} 
\frac{(r+1)(r+2)\zeta(r+2)}{\zeta(r+1)}  
\hspace{1cm}\mbox{for $ q = 1$}.
\label{eq:S6} 
\end{eqnarray}

The calculated specific heats for $r=0.5$ and 2.0 are plotted
in Figs. \ref{figH}(a) and (b), respectively.
The magnitude of the specific heat
is monotonously increased with increasing $q$ and/or $r$. 
A jump in the specific heat at $T_c$ is given by
\begin{eqnarray}
\frac{\Delta C}{k_B N} &=& \frac{C(T_c-0)-C(T_c+0)}{k_B N}, \\
&=&  \frac{(r+1)^2 \zeta(r+1)}{\zeta(r)}
\hspace{1cm}\mbox{for $0< q < 3$}.
\label{eq:S7}
\end{eqnarray}
Equation (\ref{eq:S7}) shows that for $r \leq 1.0$,
$\Delta C$ vanishes and $C$ is continuous at $T_c$  
because of the divergence in $\zeta(r)$. 
Then $\Delta C$ vanishes for $r=0.5$ in Fig. \ref{figB}(a)
while it is finite for $r=2.0$ in Fig. \ref{figB}(b).

\section{BCS supercondunctivity} 

\subsection{Model hamiltonian}

The BCS model hamiltonian is given by
\begin{eqnarray}
H &=& \sum_k \epsilon_k (n_k+n_{-k})
+\sum_{kk'} V_{kk'}b_{k'}^{\dagger} b_k,
\label{eq:J1}
\end{eqnarray}
where the operator $b_k^{\dagger}$ $(=c_k^{\dagger} c_{-k}^{\dagger})$
creates a Cooper pair in a singlet superconducting state.
The attractive interaction $V_{kk'}$,
which is the origin of the superconductivity, is assumed to be
effective only near the fermi level $\mu$,
\begin{eqnarray}
V_{kk'} &=& -V 
\hspace{1cm}\mbox{for $\vert \epsilon_k-\mu \vert, 
\vert \epsilon_{k'}-\mu \vert< \hbar \omega_D$},\\
&=& 0
\hspace{2cm}\mbox{otherwise},
\end{eqnarray} 
where $\hbar \omega_D= k_B \Theta_D$, and $\omega_D$ 
and $\Theta_D$ denote the Debye energy and temperature, respectively.
The BCS model was discussed with the use of the FA
\cite{Nunes01,Uys01}. We will examine nonextensive superconductivity, 
by using the $q$-FDD in the IA.

\subsection{Order parameter}
By using the double-time Green function method 
for the nonextensive quantum systems \cite{Rajagopal98,Lenzi99},
Numes and Mello \cite{Nunes01} derived the 
self-consistent equation for the order parameter $\Delta$ given by
\begin{eqnarray}
\Delta &=& V \rho \:\int_{0}^{\hbar \omega_D} 
\frac{\Delta}{E}[f_q(-E,\beta)-f_q(E,\beta)]\:d\epsilon,
\label{eq:J2}
\end{eqnarray}
with 
\begin{eqnarray}
E &=& \sqrt{\epsilon^2+\Delta^2},
\label{eq:J3}
\end{eqnarray}
where $\rho=\rho(\mu)$ expresses the density of states
at the fermi level.
The ground-state order parameter $\Delta_0$ is determined by
\begin{eqnarray}
\frac{1}{V \rho} &=& \:\int_{0}^{\hbar \omega_D} 
\frac{1}{\sqrt{\epsilon^2+\Delta_0^2}}\:d\epsilon.
\label{eq:J4}
\end{eqnarray}
Because the $q$-FDD at $T=0$ is the same as $f_1(\epsilon)$ \cite{Hasegawa09b},
the ground-state order parameter is identical with the BCS result for $q=1.0$:
\begin{eqnarray}
\Delta_0 &=&  (\hbar \omega_D 
+\sqrt{(\hbar \omega_D)^2+\Delta_0^2}) \:e^{-1/V \rho},
\end{eqnarray}
independently of $q$.
The critical temperature $T_c$ ($=1/k_B \beta_c$) where
$\Delta$ vanishes is given by
\begin{eqnarray}
\frac{1}{V \rho} &=& \int_{0}^{\hbar \omega_D}
\left[ \frac{1-2 f_q(\epsilon,\beta_c)}{\epsilon}\right]
\:d\epsilon.
\label{eq:J5} 
\end{eqnarray}

We show in Fig. \ref{figA}, the $q$-FDD calculated
by the IA and FA for $q=0.8$ and 1.2, the result for
$q=1.0$ being plotted by the chain curve.
For $q = 1.2$, a tail of the distribution at large $\epsilon$ 
obeys the power law.
In contrast, the distribution for $q = 0.8$ has a compact form 
with the cut-off properties:
$f_q(\epsilon)=1.0$ for $\beta (\epsilon-\mu) \leq - 5.0$
ad $f_q(\epsilon)=0.0$ for $\beta (\epsilon-\mu) \geq 5.0$.
These properties in the $q$-FDD distribution are
more clearly realized in its derivative, 
$- \partial f_q(\epsilon)/\partial \epsilon$, which is plotted in
the inset of Fig. \ref{figA}.
We note that $- \partial f_q(\epsilon)/\partial \epsilon$ in the IA
is symmetric with respect to $\epsilon=\mu$ independently of $q$,
while that in the FA is not for $q \neq 1.0$.

We have solved the self-consistent equation (\ref{eq:J2}), 
by using the $q$-FDD in the IA.
Figure \ref{figS}(a) shows the temperature dependence of
the order parameter for various $q$ values with $V \rho=0.3$.
With decreasing $q$ from unity, the temperature dependence
of the order parameter become significant and
the Curie temperature is decreased.
On the contrary, with increasing $q$ above unity, 
we observe the opposite tendency: the temperature dependence
of the order parameter become less significant and
the Curie temperature is increased.
The properties of the order parameter are clearly realized 
in Fig. \ref{figS}(b), where the normalized order parameter is plotted.
The $q$ dependence of the normalized critical temperature,
$T_{c,q}/T_{c,1}$, is plotted in
Fig. \ref{figU}, showing an monotonous increase in $T_c$ 
with increasing $q$. 
The dashed curve in Fig. \ref{figU} shows 
the $q$ dependence of the critical temperature
calculated with the $q$-FDD in the FA,
which shows the different $q$ dependence of $T_c$ from that in the IA.
The chain curve will be discussed shortly.

We may obtain the $(q-1)$ expansion
of the superconducting critical temperature,
by using the $(q-1)$ expansion of $f_q^{IA}(\epsilon)$ 
given in Table 1, leading to
\begin{eqnarray}
\frac{1-2 f_q^{IA}(\epsilon)}{x}
&=& \frac{1-2 f_1(\epsilon)}{x}
-2(q-1)\left[ \frac{\partial f_1(\epsilon)}{\partial x}
+\frac{x}{2} \: \frac{\partial^2 f_1(\epsilon)}{\partial x^2} \right]
+ \cdot\cdot.
\label{eq:H1}
\end{eqnarray}
Substituting Eq. (\ref{eq:H1}) to Eq. (\ref{eq:J5}), we obtain
\begin{eqnarray}
\frac{1}{V \rho} &=& \ln \left( \frac{\hbar \omega_D}{2 \pi T_c} \right)
- \ln \left( \frac{e^{-\gamma}}{4}\right) +\frac{1}{2}(q-1)+\cdot\cdot,
\label{eq:H2}
\end{eqnarray}
from which the critical temperature is given by
\begin{eqnarray}
T_{c,q} &=& T_{c,1} \left[1+\frac{1}{2}(q-1)+\cdot\cdot \right],
\label{eq:H3}
\end{eqnarray}
$T_{c,1}$ [=$(2 e^{\gamma}/\pi) \:\hbar \omega_D\;e^{-1/V\rho}$ ] 
denoting the critical temperature for $q=1.0$ and
$\gamma$ (=0.577) the Euler constant.
The chain curve in Fig. \ref{figU} expresses 
Eq. (\ref{eq:H3}), which is in good agreement with the solid curve
expressing the result obtained by a self-consistent calculation
of Eq. (\ref{eq:J2}).

\subsection{Specific heat}
The specific heat in the superconducting states
is given by
\begin{eqnarray}
C_s &=& \frac{4}{T} \int_{0}^{\omega_D}
\left( E^2 - \frac{T}{2}\frac{d \Delta^2}{d T} \right)
\left( \frac{- \partial f_q(E)}{\partial E} \right) \:d\epsilon,
\end{eqnarray}
while that in the normal states is given by
\begin{eqnarray}
C_n &=& \frac{4}{T} \int_{0}^{\omega_D} \epsilon^2
\left( \frac{- \partial f_q(\epsilon)}{\partial \epsilon} \right) \:d\epsilon.
\end{eqnarray}
Figure \ref{figW} shows the temperature dependence of the
specific heat for various $q$ values 
calculated with the use of $q$-FDD in the IA.
The temperature dependence of the specific heat shows 
the well-known exponential temperature dependence
in the superconducting states for $q=1.0$. 
When $q$ is decreased from unity,
this behavior becomes more significant.
On the contrary, when $q$ is increased from unity, the exponential 
temperature dependence is changed to nearly linear and then convex behavior.
There is a jump in the specific heat at $T_c$ given by
\begin{eqnarray}
\Delta C &=& C_s(T_c-0)-C_n(T_c+0), \\ 
&=& -2 \frac{d \Delta^2}{d T}
\int_0^{\omega_D} 
\left( -\frac{\partial f_q(\epsilon)}{\partial \epsilon} \right).
\end{eqnarray}

It is interesting to investigate the $q$ dependence of
the ratios given by
\begin{eqnarray}
\frac{2 \Delta_0}{k_B T_c}, \hspace{1cm}
\left( \frac{\Delta C}{C_n} \right)_{T_c},
\end{eqnarray}
which are 3.53 and 1.43, respectively, for $q=1.0$ in the BCS theory.
Solid curves in Fig. \ref{figX} express
the $q$ dependence of these ratios  
calculated in the IA, showing
that both the ratios are deceased with increasing $q$.
In contrast, the $q$ dependence of the ratios calculated 
in the FA shown by dashed curves
are rather different: with increasing $q$,
$\Delta C/C_n$ is decreased but $2 \Delta_0/k_B T_c$
is increased. The $q$ dependence in $2 \Delta_0/k_B T_c$ is
due to the $q$-dependent $T_c$ shown in Fig. \ref{figU}
because $\Delta_0$ is independent of $q$. 

Table 2 shows observed ratios of $2 \Delta_0/k_B T_c$ and
$\Delta C/C_n$ of typical superconducting materials \cite{Meservey69}:
$2 \Delta_0/k_B T_c$ vs. $\Delta C/C_n$ plot of these data is presented
in Fig. \ref{figY}.
We note the tendency such that materials with larger $2 \Delta_0/k_B T_c$
have larger $\Delta C/C_n$, which is consistent with
our calculation shown by the solid curve, except for Hg and Pb which
are considered not to be weak-coupling superconductors.
From Eq. (\ref{eq:H3}), we obtain
\begin{eqnarray}
\left( \frac{2 \Delta_0}{k_B T_{c,q}} \right) &=& 
\left( \frac{2 \Delta_0}{k_B T_{c,1}} \right)
\left[1-\frac{1}{2}(q-1)+\cdot\cdot \right].
\label{eq:H4}
\end{eqnarray}
Applying Eq. (\ref{eq:H4}) to the observed values of $2 \Delta_0/k_B T_c$, 
we have estimated $q$ values of the materials, 
which are shown in the fourth column in Table 2.
A similar estimate of $q$ values for some BCS materials 
was made in \cite{Nunes01} but using the FA.
The FA, however, yields the tendency such that materials 
with larger $2 \Delta_0/k_B T_c$ have {\it smaller} $\Delta C/C_n$, 
in contrast with our result in the IA.

\begin{table}
\begin{center}
\caption{Ratios of $2 \Delta_0/k_B T_c$ and $\Delta C/C_n$
of typical BES superconductors$^{a}$ and
estimated $q$ values with the use of Eq. (\ref{eq:H4})}
\vspace{0.5cm}
\renewcommand{\arraystretch}{1.5}
\begin{tabular}{|c|c|c|c|} \hline
materials & $2 \Delta_0/k_B T_c$  & $\Delta C/C_n$ 
& q \\ \hline \hline
(BCS)  & 3.53 & 1.43  &  1.00 \\ \hline
Cd  & 3.44 & 1.32, 1.40 &  1.05 \\ \hline
Zn  & 3.44 & 1.3  &  1.05 \\ \hline
V  & 3.50 & 1.49 &  1.01 \\ \hline
Al  & 3.53 & 1.29$-$1.59  &  1.00 \\ \hline
Sn  & 3.61 & 1.60  &  0.96 \\ \hline
Tl  & 3.63 & 1.5 &  0.94 \\ \hline
Ta  & 3.63 & 1.59 &  0.94 \\ \hline
In  & 3.65 & 1.73 &  0.93 \\ \hline
Nb  & 3.65 & 1.87  &  0.93 \\ \hline
Hg  & 3.95 & 2.37  &  0.76 \\ \hline
Pb  & 3.95 & 2.71  &  0.76 \\ \hline

\end{tabular}
\end{center}

$^a$ Ref. \cite{Meservey69}


\end{table}

\section{Itinerant-electron ferromagnets}
\subsection{The Hartree-Fock approximation}

We will discussed itinerant-electron (metallic)
ferromagnets described by the Hubbard model 
given by \cite{Hasegawa09a}\cite{Hubbard64}
\begin{eqnarray}
\hat{H} &=& \sum_{\sigma}\sum_i \epsilon_0 \:n_{i \sigma} 
+\sum_{\sigma} \sum_{i,j} t_{ij} \:a^{\dagger}_{i \sigma}a_{j \sigma}
+ U \:\sum_i n_{i\uparrow}n_{i\downarrow}
-\mu_B B \:\sum_i(n_{i\uparrow}-n_{i\downarrow}). \nonumber \\
&& \hspace{5cm}
\label{eq:B1}
\end{eqnarray}
Here $n_{i \sigma}=a^{\dagger}_{i \sigma}a_{i \sigma}$,
$a_{i \sigma}$ ($a^{\dagger}_{i \sigma}$) denotes  
an annihilation (creation) operator of 
a $\sigma$-spin electron ($\sigma=\uparrow,\:\downarrow $)
at the lattice site $i$, $\epsilon_0$ the intrinsic energy of atom,
$t_{ij}$ the electron hopping, $U$ the intra-atomic electron-electron
interaction, $B$ an applied magnetic field
and $\mu_B$ the Bohr magneton,
With the use of the Hartree-Fock approximation, 
Eq. (\ref{eq:B1}) becomes 
the effective one-electron Hamiltonian given by
\begin{eqnarray}
\hat{H} &=& \sum_{\sigma}\sum_i \epsilon_0 \:n_{i \sigma} 
+\sum_{\sigma} \sum_{i,j} t_{ij} \:a^{\dagger}_{i \sigma}a_{j \sigma}
+ U \:\sum_i
(\langle n_{i\downarrow}\rangle n_{i\uparrow}
+\langle n_{i\uparrow} \rangle n_{i\downarrow}) \nonumber \\
&& - \mu_B\: B\sum_i (n_{i \uparrow }-n_{i \downarrow}),
\label{eq:B2}
\end{eqnarray}
where the bracket $\langle \cdot \rangle $ denotes
the expectation value [Eq. (\ref{eq:C3})].
This subject was previously discussed in Ref. \cite{Hasegawa09a}
with the use of the FA.

\subsection{Magnetic moment}

Self-consistent equations
for the magnetic moment ($m$) and the number of
electrons ($n$) per lattice site are given by \cite{Hasegawa09a}
\begin{eqnarray}
m &=&  \langle n_{\uparrow} \rangle -  \langle n_{\downarrow} \rangle,
\label{eq:C1} \\
n &=& \langle n_{\uparrow} \rangle + \langle n_{\downarrow}  \rangle,
\label{eq:C2}
\end{eqnarray}
with
\begin{eqnarray}
\langle n_{\sigma} \rangle
&=& \int \:\rho_{\sigma}(\epsilon)f_{q}(\epsilon)\:d\epsilon,
\label{eq:C3} \\
\rho_{\uparrow, \downarrow}(\epsilon) 
&=& \rho_0\left( \epsilon -\epsilon_0- \frac{U}{2}(n \mp m) 
\pm \mu_B B \right),
\label{eq:C4} \\
\rho_0(\epsilon) &=& \frac{1}{N_a} \sum_k \delta(\epsilon-\epsilon_k),
\label{eq:C5}
\end{eqnarray}
where $f_q(\epsilon)$ expresses the $q$-FDD given 
by Eqs. (\ref{eq:N5})-(\ref{eq:N7}),
$\rho_0(\epsilon)$ denotes the density of states,
$\epsilon_k$ is the Fourier transform of $t_{ij}$ and
$N_a$ the number of lattice sites:
the plus and minus signs in Eq. (\ref{eq:C4}) are applied to
$ \uparrow$- and $ \downarrow$-spin electrons, respectively. 
From Eqs. (\ref{eq:C1})-(\ref{eq:C5}), $m$ and $\mu$ are 
self-consistently determined
as a function of $T$ for given parameters of
$q$, $n$ and $U$ and density of state, $\rho_0(\epsilon)$.

We have performed model calculations
bearing in mind Fe, which has seven d electrons and 
the ground-state magnetic moment of 2.2 $\mu_B$.  
By using a bell-shape density of states 
for a single band given by \cite{Hasegawa09a}
\begin{eqnarray}
\rho_0(\epsilon) &=& \left( \frac{2}{\pi W} \right)
\sqrt{1-\left(\frac{\epsilon}{W}\right)^2} 
\;\Theta(W- \vert \epsilon \vert),
\label{eq:C6}
\end{eqnarray}
we have adopted $U/W=1.75$ and $n=1.4$ electrons as in \cite{Hasegawa09a},
where $W$ denotes a half of the total bandwidth.

Fig. \ref{figB} shows the temperature 
dependence of the magnetic moment $m$ for $q=0.8$, 1.0 and 1.2
calculated by the IA and FA. 
For $q=1.2$, the temperature dependence of
magnetic moments becomes more significant and the
Curie temperature becomes lower than for $q=1.0$ in the IA. 
On the other hand, for $q=0.8$, the temperature dependence of
$m$ becomes less significant and the Curie temperature 
becomes higher than for $q=1.0$ in the IA.
The behavior of $m$ in the FA is quite different from that 
in the IA: the Curie temperature is more decreased both 
for $q=0.8$ and 1.2 than for $q=1.0$.
This fact is more clearly seen in Fig. \ref{figC}, where
$T_C$ is plotted as a function of $q$.
The Curie temperature in the IA monotonously decreased with
increasing $q$. On the contrary, $T_C$ in the FA
is almost symmetric with respect to $q=1.0$
where we obtain the maximum value of $k_B T_C/W=0.143$.
If we adopt $W \simeq 2.5$ eV obtained by the band-structure calculation 
for Fe \cite{Callaway97}, the calculated Curie temperature 
at $q=1.0$ is $T_C \simeq 3500$ K, 
while the observed $T_C$ of Fe is 
$1044$ K \cite{Crangle71}.

\subsection{Energy and Specific heat}

The energy per lattice site is expressed by \cite{Hasegawa09a}
\begin{eqnarray}
E &=& \int \epsilon 
\: [\rho_{\uparrow}(\epsilon)+\rho_{\downarrow}(\epsilon)]
f_q(\epsilon)\: d\epsilon 
- \frac{U}{4}(n^2-m^2),
\label{eq:D1}
\end{eqnarray}
from which the electronic specific heat is given by
\begin{eqnarray}
C &=& \frac{d E}{dT}=\frac{\partial E}{\partial T}
+ \frac{\partial E}{\partial m} \frac{d m}{d T}
+ \frac{\partial E}{\partial \mu} \frac{d \mu}{d T},
\label{eq:D2}
\end{eqnarray}
with
\begin{eqnarray}
\frac{\partial E}{\partial T}
&=& - \frac{1}{T} \int \epsilon\:(\epsilon-\mu)
[\rho_{\uparrow }(\epsilon)+\rho_{\downarrow}(\epsilon)] 
\:\frac{\partial f_q(\epsilon)}{\partial \epsilon} \: d\epsilon, 
\label{eq:D3} \\
\frac{\partial E}{\partial m}
&=&- \frac{U}{2} \int \epsilon
\:[\rho_{\uparrow }(\epsilon)-\rho_{\downarrow}(\epsilon)] 
\:\frac{\partial f_q(\epsilon)}{\partial \epsilon} 
\: d\epsilon, 
\label{eq:D4} \\
\frac{\partial E}{\partial \mu}
&=& -\int \epsilon
\: [\rho_{\uparrow }(\epsilon)+\rho_{\downarrow}(\epsilon)] 
\:\frac{\partial f_q(\epsilon)}{\partial \epsilon} 
\: d\epsilon.
\label{eq:D5}
\end{eqnarray}
Analytic expressions for $d m/dT$ and $d\mu/dT$ in Eq. (\ref{eq:D2}) 
are given by Eqs. (A.3)-(A.8) in Ref. \cite{Hasegawa09a}. 

Fig. \ref{figD} shows the temperature 
dependence of the specific heat $C$ for $q=0.8$, 1.0 and 1.2,
calculated by using the IA and FA.
In the IA, $C$ for $q=0.8$ is smaller than that for $q=1.0$.
In contrast, $C$ in the FA of $q=0.8$ is larger than that of $q=1.0$. 

\subsection{Spin susceptibility}

The spin susceptibility is expressed by \cite{Hasegawa09a}
\begin{eqnarray}
\chi &=& \frac{d m}{d B}, 
\label{eq:E1}
\end{eqnarray}
from which the paramagnetic spin susceptibility is given by
\begin{eqnarray}
\chi &=& \mu_B^2 \:\frac{2 \chi_{0}}{(1-U \chi_{0})},
\label{eq:E2}
\end{eqnarray}
with
\begin{eqnarray}
\chi_0 &=& -\int  \: \rho(\epsilon) 
\frac{\partial f_q(\epsilon)}{\partial \epsilon}\:d\epsilon.
\label{eq:E3}
\end{eqnarray}

Figs. \ref{figE} shows the temperature dependence of the inversed
susceptibility $1/\chi$ calculated by the IA and FA
for $q=0.8$, 1.0 and 1.2.
The Curie temperature $T_C$, which is realized at $1/\chi=0$,
is monotonously decreased with increasing $q$ in the IA, 
which is different from its $q$ dependence in the FA,
as shown in Fig. \ref{figC}.

\section{Discussion}

We will qualitatively elucidate the
difference between the results calculated with the IA and FA
for itinerant-electron ferromagnets. 
The generalized Sommerfeld expansion 
including an arbitrary function $\phi(\epsilon)$
and the $q$-FDD $f_q(\epsilon)$ 
is given by \cite{Hasegawa09b,Hasegawa09a}
\begin{eqnarray}
I &=& \int  \: \phi(\epsilon) f_q(\epsilon)\: d\epsilon, \\
&=& \int^{\mu}  \: \phi(\epsilon)\:d\epsilon
+ \sum_{n=1}^{\infty} \:c_{n,q} \:(k_B T)^n \phi^{(n-1)}(\mu),
\label{eq:F1}
\end{eqnarray}
with
\begin{eqnarray}
c_{n,q} &=& - \frac{\beta^n}{n!} \int
(\epsilon-\mu)^n \: \frac{\partial f_q(\epsilon)}{\partial \epsilon}
\:d \epsilon,
\label{eq:F2}
\end{eqnarray} 
which is valid at low temperatures. 
Expansion coefficients for $q=1.0$
are given by $c_{2,1}=\pi^2/6$ (=1.645), 
$c_{4,1}=7 \pi^4/360$ (=1.894), 
and $c_{n,1}=0.0$ for odd $n$.
The coefficients $c_{n,q}$ for $n=2$ and 4 in the IA
are given by \cite{Hasegawa09b}
\begin{eqnarray}
\frac{c_{2,q}^{IA}}{c_{2,1}} &=& 
\frac{1}{(2-q)}, 
\label{eq:F3} \\ 
\frac{c_{4,q}^{IA}}{c_{4,1}} &=& 
\frac{1}{(2-q)(3-2q)(4-3q)},
\label{eq:F4}
\end{eqnarray}
whereas $c_{1,q}^{IA}=c_{3,q}^{IA}=0$.
Results in the IA is in agreement with those of the EA within $O(q-1)$
\cite{Hasegawa09b}.

On the other hand, the FA yields \cite{Hasegawa09a}
\begin{eqnarray}
c_{1,q}^{FA} &=& \frac{\pi^2}{6}(q-1) + \cdot\cdot, \nonumber \\
c_{2,q}^{FA} &=& \frac{\pi^2}{6}[1  + a (q-1)^2+ \cdot\cdot], \nonumber \\
c_{3,q}^{FA} &=& \frac{7\pi^4}{60} (q-1)+ \cdot\cdot, \nonumber \\
c_{4,q}^{FA} &=& \frac{7 \pi^4}{360}[1 + b (q-1)^2+ \cdot\cdot], \nonumber
\end{eqnarray}
where $a \simeq 9.2$, $b \simeq 110$ and
the $O(q-1)$ contributions to $c_{2,q}^{FA}$ and $c_{4,q}^{FA}$
are vanishing \cite{Hasegawa09a}.
$c_{1,q}^{FA}$ and $c_{3,q}^{FA}$
are not zero, which is in contrast with the results of the EA and IA.
This is due to a lack of the symmetry 
in $-\partial f_q^{FA}(\epsilon)/\partial \epsilon$,
as shown in the inset of Fig. \ref{figA}.

Figure \ref{figF} shows the $q$ dependence of 
$c_{n,q}$ for $n=1-4$ calculated by the IA and FA.
We note that the $q$ dependence of $c_{2,q}^{FA}$ and $c_{4,q}^{FA}$
is symmetric with respect to $q=1.0$ 
whereas that in the IA is not.

By simple calculations using 
Eqs. (\ref{eq:C1}), (\ref{eq:C2}), (\ref{eq:D1}), (\ref{eq:E2}),
(\ref{eq:F1}) and  (\ref{eq:F2}),
we obtain the magnetic moment $m(T)$, the specific
heat $C$ at low temperatures and the Curie temperature
$T_{C,q}$ given by \cite{Hasegawa09a}
\begin{eqnarray}
m(T) &\simeq & m(0) 
-\alpha \:T^2, 
\label{eq:G1} \\
C(T) &\simeq & \gamma_q \:T, 
\label{eq:G2} \\
T_{C,q}&\simeq& \left( \frac{U\rho-1}
{-c_{2,q} \rho^{(2)}} \right)^{1/2}, 
\label{eq:G3}
\end{eqnarray}
with
\begin{eqnarray}
\alpha &=& c_{2,q} \:(\rho'_{\downarrow}- \rho'_{\uparrow}),
\label{eq:G4} \\
\gamma_q &=& 2c_{2,q}[2(\rho_{\uparrow}+\rho_{\downarrow})
-U m(0) (\rho_{\uparrow}-\rho_{\downarrow})], 
\label{eq:G5}
\end{eqnarray} 
where $\rho_{\sigma}=\rho_{\sigma}(\mu)$, $\rho=\rho(\mu)$, 
$\rho^{(2)}=d^2 \rho(\mu)/d \epsilon^2$,
and $m(0)$ is the
ground-state magnetic moment.  
Equation (\ref{eq:G3}) 
leads to
\begin{eqnarray}
\frac{T_{C,q}}{T_{C,1}} 
&\simeq & \left(\frac{c_{2,q}}{c_{2,1}}\right)^{-1/2},\\
&\simeq& 1 -\frac{(q-1)}{2}+\cdot\cdot 
\hspace{1cm}\mbox{in the IA}, 
\label{eq:G6} \\
&\simeq& 1 -\frac{a (q-1)^2}{2}+\cdot\cdot 
\hspace{1cm}\mbox{in the FA}.
\label{eq:G7}
\end{eqnarray}
Equation (\ref{eq:G6}) shows that
with increasing (decreasing) $q$ from unity, the Curie temperature in the IA is 
decreased (increased). 
In contrast, Eq. (\ref{eq:G7}) shows that
the Curie temperature in the FA is decreased
with increasing $\vert q-1 \vert$.
These are
consistent with the results shown in Fig. \ref{figC}. 
The coefficient of $c_{2,q}$ expresses the contribution from the Stoner excitations, 
which play important roles in magnetic and thermodynamical 
properties of itinerant-electron ferromagnets.
It is noted from Eqs. (\ref{eq:R5}) and (\ref{eq:F4}) that 
the Stefan-Boltzmann constant $\sigma_q$ is related to the 
Sommerfeld expansion coefficient $c_{4,q}$ as given by
$\sigma_q/\sigma_1=c_{4,q}/c_{4,1}$.
Thus the difference in the Sommerfeld expansion coefficients in the IA and FA
reflects on the difference in the $q$ dependence of the
physical quantities calculated by the two kinds of approximations.

\section{Conclusion}

By using the generalized distributions of $q$-BED and $q$-FDD 
in the IA which was proposed in Ref. \cite{Hasegawa09b},
we have discussed four nonextensive quantum subjects: 
(1) the black-body radiation, (2) Bose-Einstein condensation,
(3) the BCS superconductivity and (4) itinerant-electron ferromagnetism. 
The effect of the nonextensivity of $(q-1)$ has been shown to be appreciable
in these systems. 
A comparison between the results obtained by the IA and FA has shown that
the FA generally overestimates the effect of the nonextesivity of $\vert q-1 \vert$
and that the $q$-FDD in the FA may yield qualitatively inaccurate results
for fermion systems such as (3) and (4).
This is considered to be due to the inappropriate $q$-FDD in the FA.
It would be necessary to examine the validity 
of the previous studies which have made with the use of the FA.

\section*{Acknowledgments}
This work is partly supported by
a Grant-in-Aid for Scientific Research from the Japanese 
Ministry of Education, Culture, Sports, Science and Technology.  



\newpage

\begin{figure}
\begin{center}
\end{center}
\caption{
(Color online)
The $\omega$ dependence of (a) $f_q(\hbar \omega)$ and
(b) $D_q(\omega)$ calculated by the IA (solid curves)
and FA (dashed curves) for $q=0.8$ and 1.2,
result for $q=1.0$ being plotted by chain curves.
}
\label{figL}
\end{figure}

\begin{figure}
\begin{center}
\end{center}
\caption{
(Color online)
The $q$ dependence of the coefficient
of the generalized Stefan-Boltzmann law, $\sigma_q$, calculated
in the IA (the solid curve) and FA (the dashed curve).
}
\label{figK}
\end{figure}

\begin{figure}
\begin{center}
\end{center}
\caption{
(Color online)
The $q$ dependence of 
the generalized Wien shift law, $\omega_{m,q}$,
calculated in the IA (the solid curve), FA (the dashed curve)
and AA (the chain curve) \cite{Tsallis95}.
}
\label{figJ}
\end{figure}

\begin{figure}
\begin{center}
\end{center}
\caption{
(Color online)
The $q$ dependence of  the critical temperature of the 
Bose-Einstein condensation,
$T_{c,q}$, for (a) $r=0.5$ and (b) $r=2.0$
calculated by the IA (solid curves) and FA (dashed curves).
}
\label{figG}
\end{figure}

\begin{figure}
\begin{center}
\end{center}
\caption{
(Color online)
The temperature dependence of the specific heat $C$
for (a) $r=0.5$ and (b) $r=2.0$
calculated in the IA, $T_c$ denoting the critical temperature
for the BEC.
}
\label{figH}
\end{figure}

\begin{figure}
\begin{center}
\end{center}
\caption{
(Color online)
The $\epsilon$ dependences of the $q$-FDDs of $f_q(\epsilon)$
[$-\partial f_q(\epsilon)/\partial \epsilon$ in the inset]
for $q=0.8$ (solid curves) and 1.2 (bold solid curves) in the IA, 
and those for $q=0.8$ (dashed curves) and 1.2 (bold dashed curves) in the FA,
results for $q=1.0$ being plotted by chain curves. 
}
\label{figA}
\end{figure}

\begin{figure}
\begin{center}
\end{center}
\caption{
(Color online)
(a) The temperature dependence of the superconducting order parameter $\Delta$, 
and (b) the temperature dependence of the normalized order parameter $\Delta$ 
calculated by the IA for various $q$ values ($V \rho=0.3$).
}
\label{figS}
\end{figure}

\begin{figure}
\begin{center}
\end{center}
\caption{
(Color online)
The $q$ dependence of the critical temperature $T_{c,q}/T_{c,1}$ 
calculated by the IA (the solid curve) and FA (the dashed curve)
with $V \rho=0.3$, the chain curve denoting the result
of $(q-1)$ expansion given by Eq. (\ref{eq:H3}).
}
\label{figU}
\end{figure}

\begin{figure}
\begin{center}
\end{center}
\caption{
(Color online)
The temperature dependence of the specific heat $\Delta$ 
calculated by the IA for various $q$ values ($V \rho=0.3$).
}
\label{figW}
\end{figure}

\begin{figure}
\begin{center}
\end{center}
\caption{
(Color online)
The $q$ dependence of the ratios of $2 \Delta_0/k_B T_c$
and $\Delta C/C_n(T_c)$ calculated 
by the IA (solid curves) and FA (dashed curves)
with $V \rho=0.3$.
}
\label{figX}
\end{figure}

\begin{figure}
\begin{center}
\end{center}
\caption{
(Color online)
$2 \Delta_0/k_B T_c$ vs. $\Delta C/C_n$ for typical BCS
superconductiors (squares) with the result calculated in the IA
(the solid curve).
}
\label{figY}
\end{figure}

\begin{figure}
\begin{center}
\end{center}
\caption{
(Color online)
The temperature dependence of the magnetic moment $m$ calculated in the IA
(solid curves) and FA (dashed curves) for $q=0.8$ and $q=1.2$:
the result for $q=1.0$ is plotted by the chain curve.
}
\label{figB}
\end{figure}

\begin{figure}
\begin{center}
\end{center}
\caption{
(Color online)
The $q$ dependence of the Curie temperature $T_C$ calculated by 
the IA (the solid curve) and FA (the dashed curve).
}
\label{figC}
\end{figure}

\begin{figure}
\begin{center}
\end{center}
\caption{
(Color online)
The temperature dependence of the electronic specific heat $C$ 
calculated in the IA (solid curves) and FA (dashed curves) 
for $q=0.8$ and $q=1.2$: the result for $q=1.0$ is plotted by the chain curve.
}
\label{figD}
\end{figure}

\begin{figure}
\begin{center}
\end{center}
\caption{
(Color online)
The temperature dependence of the inverse spin susceptibility $1/\chi$ 
calculated in the IA (solid curves) and FA (dashed curves) for $q=0.8$ and $q=1.2$:
the result for $q=1.0$ is plotted by the chain curve.
}
\label{figE}
\end{figure}

\begin{figure}
\begin{center}
\end{center}
\caption{
(Color online)
The $q$ dependence of the coefficients $c_{n,q}$ for $n=1-4$ 
in the generalized Sommerfeld expansion calculated by the IA (solid curves)
and FA (dashed curves): note that $c_{1,q}=c_{3,q}=0$ in the IA
whereas $c_{1,q}\neq 0$ and $c_{3,q} \neq 0$ in the FA.
}
\label{figF}
\end{figure}

\end{document}